\documentclass[sigconf]{acmart}
\usepackage{confcopyright} 

\newtheorem{chall}{Challenge}

\AtBeginDocument{%
  \providecommand\BibTeX{{%
    \normalfont B\kern-0.5em{\scshape i\kern-0.25em b}\kern-0.8em\TeX}}}

\copyrightyear{2020} 
\acmYear{2020} 
\setcopyright{acmlicensed}\acmConference[ICSEW'20]{IEEE/ACM 42nd International Conference on Software Engineering Workshops }{May 23--29, 2020}{Seoul, Republic of Korea}
\acmBooktitle{IEEE/ACM 42nd International Conference on Software Engineering Workshops (ICSEW'20), May 23--29, 2020, Seoul, Republic of Korea}
\acmPrice{15.00}
\acmDOI{10.1145/3387940.3391469}
\acmISBN{978-1-4503-7963-2/20/05}

\begin{document}

\title[Holy Grail of Quantum Artificial Intelligence]{The Holy Grail of Quantum Artificial Intelligence: \\ Major Challenges in Accelerating the Machine Learning Pipeline}

\author{Thomas Gabor$^1$, Leo S\"unkel$^1$, Fabian Ritz$^1$, Thomy Phan$^1$, Lenz Belzner$^2$, Christoph Roch$^1$,}
\author{Sebastian Feld$^1$, Claudia Linnhoff-Popien$^1$}
\email{thomas.gabor@ifi.lmu.de}
\affiliation{%
  \institution{$^1$LMU Munich}
  \institution{$^2$MaibornWolff}
}

\renewcommand{\shortauthors}{Gabor et al.}

\begin{abstract}
  We discuss the synergetic connection between quantum computing and artificial intelligence. After surveying current approaches to quantum artificial intelligence and relating them to a formal model for machine learning processes, we deduce four major challenges for the future of quantum artificial intelligence: (i) Replace iterative training with faster quantum algorithms, (ii) distill the experience of larger amounts of data into the training process, (iii) allow quantum and classical components to be easily combined and exchanged, and (iv) build tools to thoroughly analyze whether observed benefits really stem from quantum properties of the algorithm.
\end{abstract}

\ACMCopyrightPreprint{\textit{the 1st International Workshop on Quantum Software Engineering (Q-SE)} at \textit{ICSE 2020}} 



\maketitle

\section{Motivation}

Two frontiers of research in computer science meet in the field of quantum artificial intelligence (QAI). As both artificial intelligence (AI) and quantum computing (QC) are very active fields with an overwhelming speed of new developments just within the last year~\cite{arute2019quantum,alphastar2019}, there exist vast possibilities for interactions. However, we argue that there are grand challenges which can guide us towards a  fruitful convergence of these fields.

From here on, we take a software engineer's perspective and first analyze how traditional software engineering techniques are coping with the dynamic world of AI algorithms (Section~\ref{sec:se-ai}). We proceed by pointing out which big challenges AI is going to have to face and what that might mean for the field (Section~\ref{sec:com-con}). Then we provide a quick overview of what QAI is already doing (Section~\ref{sec:qai}) and deduce the challenges revealing what QAI might mean for the general development of AI (Section~\ref{sec:chall}). We close with a very brief outlook (Section~\ref{sec:outlook}).

\section{Software Engineering for Artificial Intelligence}
\label{sec:se-ai}

\begin{figure}
	\centering
	\includegraphics[width=.72\columnwidth]{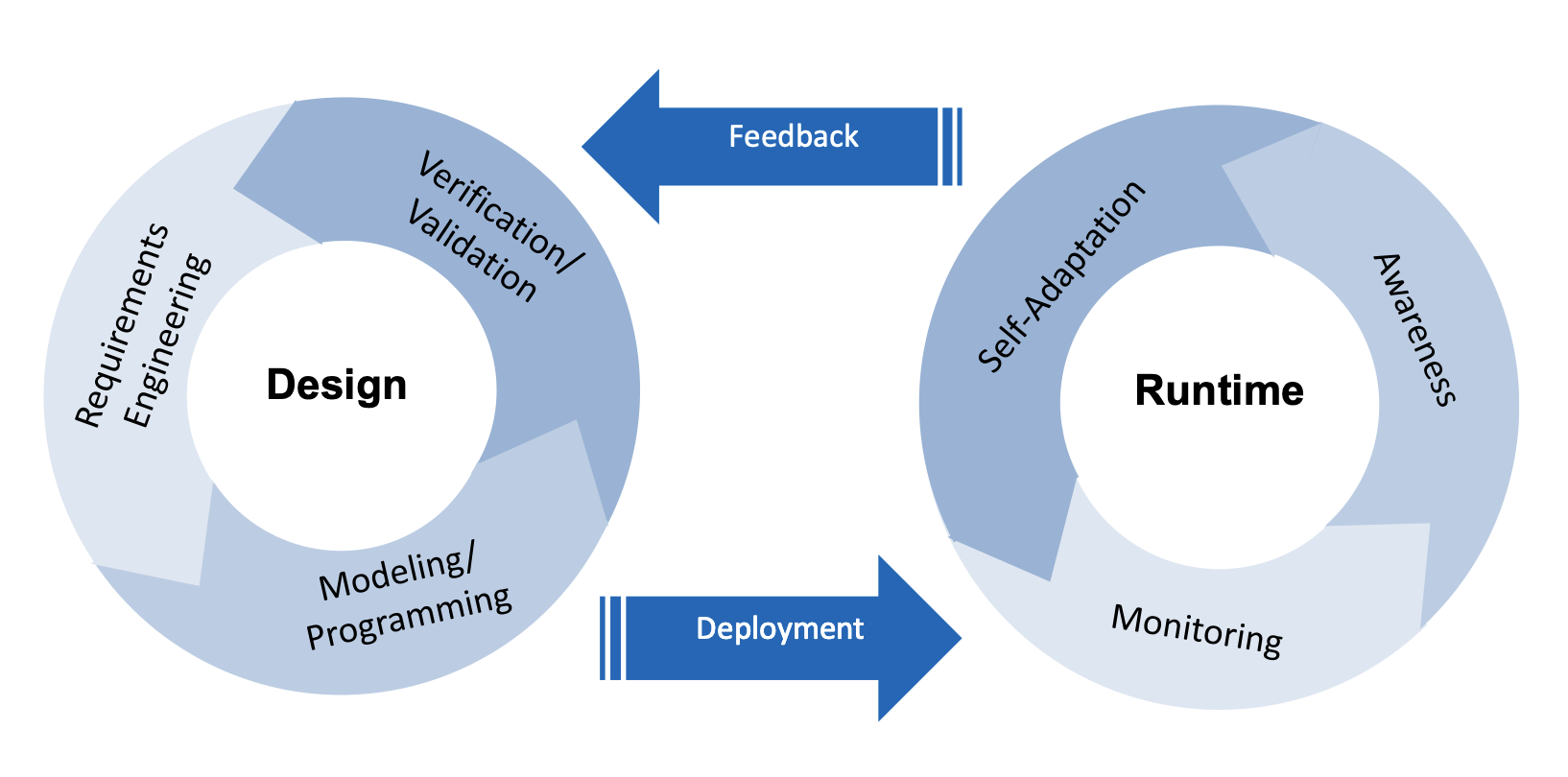}
	\caption{The ensemble development life cycle (image taken from~\cite{holzl2015ensemble}) provides a framework for the integration of design time and run-time evolution of software systems.}
	\label{fig:edlc}
	\vspace{-2em}
\end{figure}

Artificial intelligence on its own has been difficult to grasp for the art of software engineering, perhaps because traditional software engineering is focusing on preserving initial consistency (i.e., making sure the produced artifacts adhere to prior specifications)~\cite{gabor2018adapting} while methods of artificial intelligence usually start from highly chaotic initial configurations~\cite{albers1996dynamical} and only gradually introduce rules and structure. On the path towards applying the principles of rigorous engineering to more complex, adaptive and inherently self-governed systems, various directions of research have been proposed and tried~(see \cite{nierstrasz2008change,weyns2017software,bures2017software,reocas2020} and many others). As an example for these approaches, consider Figure~\ref{fig:edlc}: Classical methods of software engineering are kept as a feedback loop being driven forward usually by human developers while new ways to evolve the system at run-time are added as another feedback loop, usually driven by self-adaption and learning~\cite{holzl2015ensemble}.

\begin{figure*}
	\centering
	\includegraphics[width=.72\textwidth]{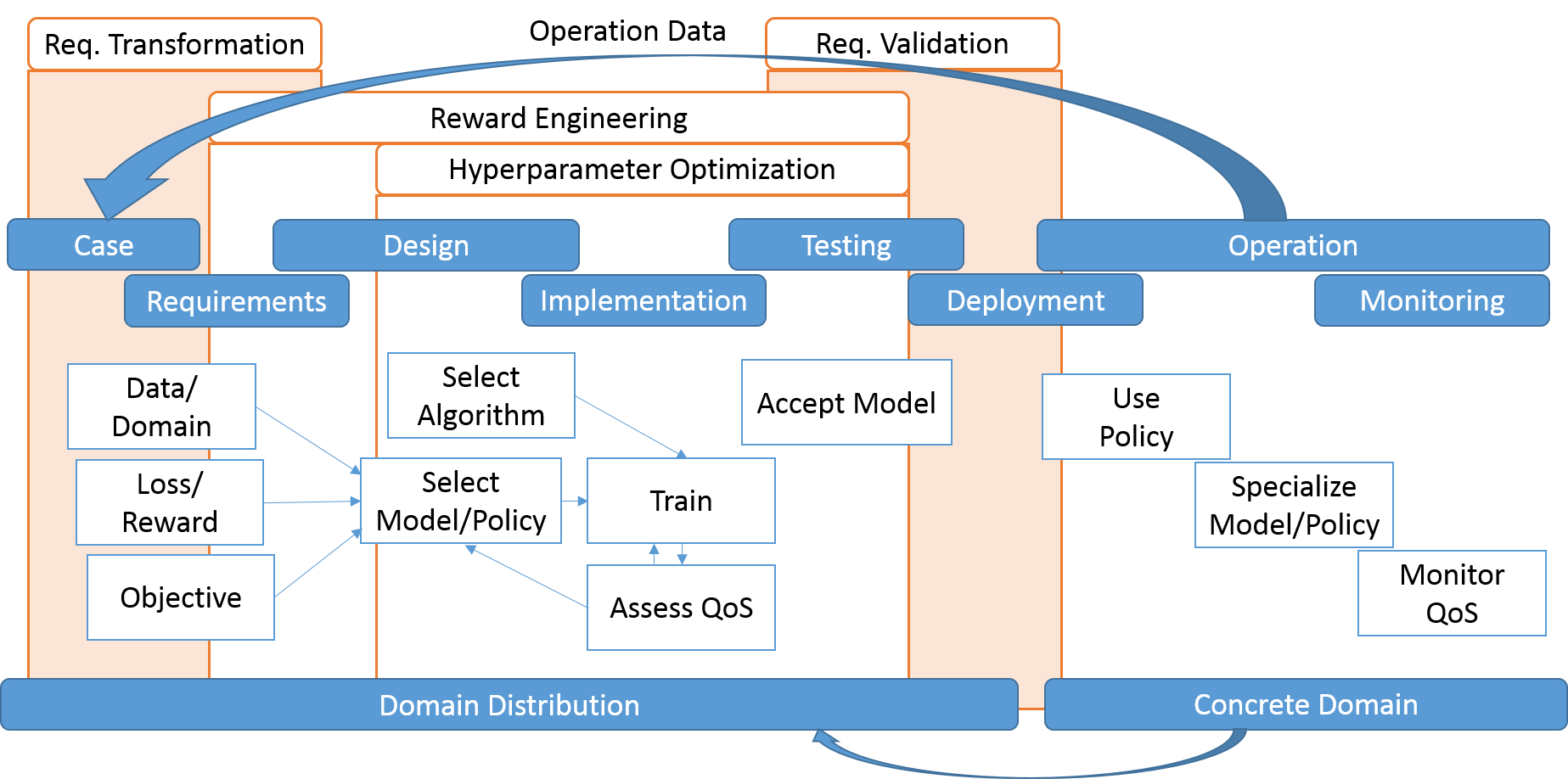}
	\caption{The machine learning pipeline (image taken from~\cite{reocas2020}).}
	\label{fig:pipeline}
\end{figure*}

However, there is a wide variety of algorithms allowing for self-adaptation and learning, ranging from simple statistical methods like SVMs or clustering to deep neural networks, and the exact way to integrate these algorithms is also subject to a lot of variation. In~\cite{reocas2020} we introduced the \emph{machine learning pipeline} as a process model for many different machine learning methods, i.e., it is a model of temporal dependences between the creation of fundamental artifacts common to most machine learning models. Figure~\ref{fig:pipeline} shows a diagram of the different tasks that are relevant to software engineering. We aim to adopt various familiar development phases from classical software engineering (blue boxes at the top level). For software engineering, there is a distinct shift between writing the software for a variety of concrete domains and specializing (a single branch of) the software to a concrete environment (blue boxes at the bottom level). The main engineering tasks are shown in white boxes. A source of great difficulty for engineering lies in the inherent stochasticity of the behavior generated by most machine learning algorithms, requiring different methods to ensure quality of service (QoS) in the main training feedback loop (``select model/policy'', ``train'', ``assess QoS'') and actual monitoring during operations. The break from most classical software engineering approaches happens here insofar we explicitly want to achieve ``softer'' behavior guidelines on the algorithms because we want to employ them in domains where we cannot possibly formulate enough ``hard'' rules. And as we still want specific behavior, the softening does not occur as random noise but is often very specific as well. Of course, this also often makes machine learning methods susceptible to systematic failures. For instance: A car driving software failing at random on one in every 100 turns is rather easy to care for by adding redundancy and a voting system, e.g., allowing us to achieve arbitrarily low overall error rates as long as we can add enough redundancy. A car driving software that operates well but systematically breaks down every time it comes across a football on the street is harder to handle because we need very specific tests to even detect the error and then see every redundant system fail at the same time.

This inherent stochasticity in machine learning algorithms, of course, is not quite unlike the inherent stochasticity in quantum computing, which is a connection already discussed (and elaborated) by  Wolfram~\cite{wolfram2018making}. For us, this may suggest that we can use similar methods to integrate (especially highly error-prone, early) quantum algorithms into classical software as we can use to integrate  the highly stochastic process called machine learning algorithms.

\section{The Role of Compute and the Consequences}
\label{sec:com-con}

Aside from similar external properties like stochasticity, quantum algorithms and artificial intelligence may indeed form an even stronger connection. The emerging field of \emph{quantum artificial intelligence (QAI)} uses quantum algorithms or quantum-inspired algorithms to solve computation tasks related to artificial intelligence.\footnote{Note that the combination also works the other way around, using AI methods to better approximate quantum computations~(for instance in~ \cite{fosel2018reinforcement,porotti2019coherent}). This, however, is beyond the scope of this paper.} This combination may be highly synergetic for two main reasons:

\begin{itemize}
	\item All machine learning methods need some randomness to work, often putting serious effort into generating necessary entropy. Beyond that, they also often show high tolerance for noise during their evolution. This makes them inherently suitable for early applications using only NISQ hardware.
	\item Progress in artificial intelligence is becoming more and more demanding in computational resources. This trend is outgrowing the continued increase in available computing power by a large margin.
\end{itemize}

The first reason basically falls in line with our point on stochasticity made earlier. While high noise levels (as they are present in NISQ machines) are unwanted for many algorithms, especially AI algorithms may actually benefit from (some levels of) noise. Of course, current noise levels over a long series of computations are way too high to even allow for meaningful results, but requirements for QAI algorithms might be met earlier than for (for example) Grover's search on similarly large input spaces.

\begin{figure}
	\centering
	\includegraphics[width=.95\columnwidth]{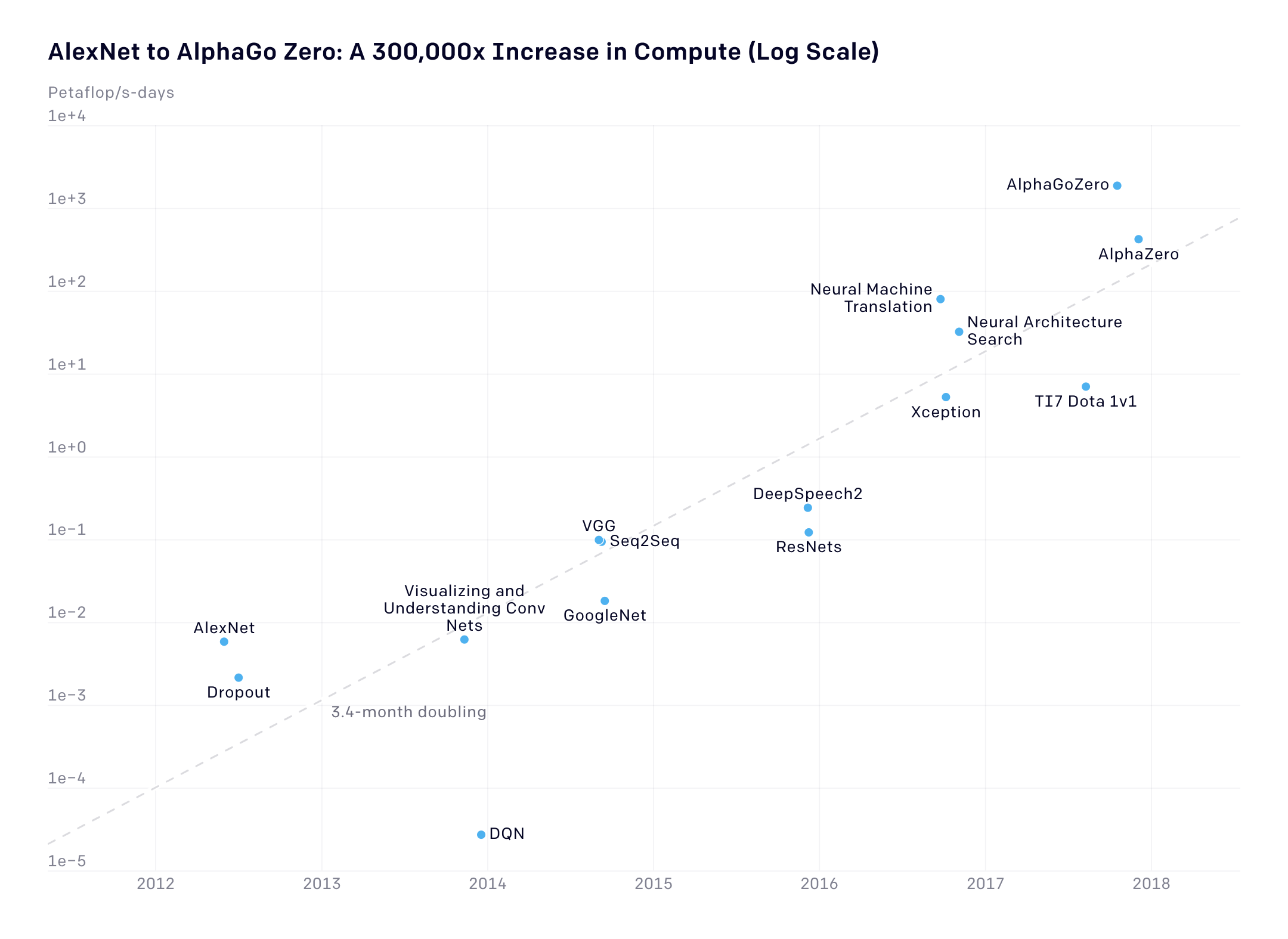}
	\caption{Computation power used for recent breakthroughs in AI over time shows a fast exponential growth (image taken from~\cite{openai2018compute}).}
	\label{fig:openai}
\end{figure}

The second reason may be a bit more elusive; of course, more computational power is always better. However, pushing the borders of AI has been especially hungry for resources. Amodei and Hernandez~\cite{openai2018compute} used the chart shown in Figure~\ref{fig:openai} to demonstrate that just in recent years, the computation power used for AI breakthrough had a doubling time of 3.5 months and has thus been dramatically outgrowing Moore's Law (18 month doubling time).

For the future of AI, this directly leads to four possible consequences (or any combination thereof):
\begin{enumerate}
	\item Progress in AI research slows down.
	\item AI research becomes exponentially more expensive.
	\item New AI algorithms using less resources are developed.
	\item New sources of computation power are discovered.
\end{enumerate}

While Consequence~1 is not entirely unlikely, it is probably the better option from a scientific point of view not to strive for that direction. Similarly, the extent to which exponentially more money for AI research (i.e., Consequence~2) can compensate the lack of computation power per chip is quite limited as we are facing an exponential demand to be satisfied.

Consequence~3 definitely should be sought for, however. Making AI algorithms more resource-efficient is imperative for many practical applications and is a rather lively topic of research~\cite{nachum2018data,cuccu2019playing}. Most interestingly, this puts AI algorithms, again, in a similar position like quantum algorithms nowadays, where working around limited hardware (albeit at an entirely different scale) is one of the key skills on bringing software to practice. However, leveraging raw computational power, i.e., using the method \emph{compute}, will probably always be a large part of using AI methods and possibly should be, as Sutton recently argued in an influential blog post~\cite{sutton2019bitter}.

Lastly, Consequence~4 suggests that new hardware might mitigate the increasing need for computational power. For some time now, we have seen this idea being implemented by using graphic cards and even more specialized hardware like neuromorphic chips to run neural networks for AI. And while substantial benefits can be achieved, none of these hardware platforms can provide an exponential speedup that can sustainably satisfy the exponential hunger of AI; but quantum computing might~\cite{arute2019quantum}.

\section{Overview of Quantum-Assisted Artificial Intelligence}
\label{sec:qai}

In order to assess the current possibilities of QAI, we performed a survey on available QAI algorithms in scientific literature, out of which we present a selection of key algorithms. For an overview, please refer to Table~\ref{table:algorithm_table}. 
We roughly identified four interesting areas of application for QAI algorithms:

\begin{enumerate}
	\item Mathematical operations. These algorithms provide faster solutions to computationally hard problems like computing eigenvalues (variational quantum eigensolver~\cite{peruzzo2014variational}) or solving linear equations (HHL~\cite{harrow2009quantum}). While these tasks are not traditionally part of machine learning, they form the basis for many models and operations used in machine learning. Any practical acceleration on that front might thus have a huge impact on QAI.
	\item Traditional machine learning. These approaches are based on methods from more traditional branches of machine learning. They train models like clusters or support vector machines (SVMs) to gain or extrapolate information from given data sets. As these models have been developed (and run) a few decades ago they are usually not as computationally expensive as other approaches of AI and might thus be fit to test QAI on limited machines.
	\item Optimization. Algorithms of this group are given a specific optimization problem (usually formulated as a goal or fitness function over a specific input space) and aim to return the globally optimal point in the input space. Purely quantum methods differ from classical stochastic optimization in that they are usually guaranteed to find the global optimum under ideal conditions. In real-world implementations, they, too, yield stochastic results. The quantum approximate optimization algorithm (QAOA)~\cite{farhi2014quantum} is able to optimize on gate model hardware. Quadratic unconstrained binary optimization (QUBO) is equivalent to Ising spin glasses and both are the canonical optimization problems for commercially available quantum annealers. They are a specific hardware platform designed to perform quantum annealing~\cite{kadowaki1998quantum}, which is a stochastic optimization algorithm based on adiabatic quantum computing (the exact ideal-condition algorithm)~\cite{mcgeoch2014adiabatic}.
	\item Neural machine learning. These algorithms are based on more modern concepts of machine learning, which usually use neural networks for models; that means their models are quite intransparent (i.e., hard to verify) and vastly over-parametrized. Some approaches opt for Boltzmann machines (BMs)~\cite{amin2018quantum,wiebe2019generative} as a model representation (since especially restricted BMs are easier to train). Others incorporate quantum computers in just select parts of larger neural-network-based architectures like Autoencoders~\cite{romero2017quantum,khoshaman2018quantum}, Generative Adversarial Networks (GANs)~\cite{dallaire2018quantum,lloyd2018quantum,romero2019variational,zoufal2019quantum}, or reinforcement learning (RL) agents~\cite{neukart2018quantum,dong2008quantum}. Here, Quantum RL~\cite{dong2008quantum} is especially interesting since it largely differs from classical RL and substitutes various artifacts and operations with their quantum analogue.
\end{enumerate}

In Table~\ref{table:algorithm_table} we also annotated all described algorithms with the QC platform they are run on as well as direct references to implementations where they were available. We attempted to guess when some algorithms will be ready for practical use based on their fitness to be run on NISQ devices. However, not for all algorithms a valid estimate was possible yet. If predictions are right, NISQ-ready algorithms will probably be the first ones to make an impact in real-world QAI. Lastly, we matched all QAI algorithms to the machine learning pipeline (cf. Figure~\ref{fig:pipeline}) and denoted the tasks of the machine learning pipeline that were actually executed using quantum computing. This usually meant that all other tasks are still performed on classical hardware.

Even in this small sample set of algorithms, we can notice that quantum algorithms can be used in various places throughout the machine learning pipeline. Naturally, they are focused on computationally expensive tasks. These can mainly be found when modeling the domain, which often means modeling complex probability distributions and sampling from them, and when performing the training, which usually means executing rather lengthy update operations on matrices or similar data structures. However, we can also observe that QAI algorithms naturally are hybrid approaches: A lot of classical steps within the machine learning pipeline are still necessary to produce the results.

\begin{table*}[h]
\footnotesize
\begin{tabular}{| l | c | c | c | c | }
\hline
\textbf{Algorithm/Task} & \textbf{QC platform} & \textbf{Impl. available} & \textbf{NISQ} & \textbf{Quantum tasks in ML pipeline} \\ \hline
\hline
Variational quantum eigensolver \cite{peruzzo2014variational} & Gate model & PennyLane \cite{vqa_pennylane} & Yes & Data/Domain, Use Policy \\ \hline 
HHL \cite{harrow2009quantum} & Gate model & Qiskit \cite{hhl_qiskit} & Unlikely \cite{preskill2018quantum} & Data/Domain, Train \\ \hline 
\hline
Clustering \cite{aimeur2007quantum} & Gate model & - & No? & Data/Domain, Use Policy \\ \hline 
Clustering \cite{lloyd2013quantum} & Gate model & - & Yes? & Data/Domain, Use Policy  \\ \hline 
Quantum nearest-neighbor \cite{wiebe2014quantum_algorithms} & Gate model & - & - & Data/Domain, Use Policy \\ \hline 
Recommendation system \cite{kerenidis2016quantum} & Gate model & - & Unlikely \cite{preskill2018quantum} & Data/Domain, Use Policy \\ \hline 
SVM \cite{havlivcek2018supervised} & Gate model & Qiskit \cite{svm_qiskit} & Yes & Data/Domain, Use Policy \\ \hline 
SVM \cite{willsch2019support} & Quantum annealing & - & - & Data/Domain, Use Policy \\ \hline 
\hline
QAOA \cite{farhi2014quantum} & Gate model & PennyLane \cite{qaoa_pennylane} & Yes & Train \\ \hline 
QUBO / Ising spin glasses~\cite{glover2018tutorial,lucas2014ising} & Quantum annealing & D-WAVE~\cite{mcgeoch2013experimental} & Yes & Train \\ \hline
Quantum-assisted EA \cite{king2019quantum} & Quantum annealing & - & - & Train \\ \hline 
\hline
Quantum BM \cite{wiebe2019generative} & Gate model & - & Yes & Train \\ \hline 
Quantum BM \cite{amin2018quantum} & Quantum annealing & - & - & Train \\ \hline 
Autoencoder \cite{romero2017quantum} & Gate model & \cite{autoencoder_implementation} & Yes & Train \\ \hline 
Autoencoder \cite{khoshaman2018quantum} & Quantum annealing & - & - & Train \\ \hline 
Quantum GAN \cite{dallaire2018quantum, lloyd2018quantum} & Gate model & PennyLane \cite{qgan_pennylane} & Yes & Data/Domain \\ \hline 
Quantum GAN \cite{romero2019variational} & Gate model & - & Yes & Data/Domain \\ \hline 
Quantum GAN \cite{zoufal2019quantum} & Gate model & Qiskit \cite{qgan_qiskit} & Yes & Data/Domain \\ \hline 
Quantum-enhanced RL \cite{neukart2018quantum} & Quantum annealing & - & - & Train \\ \hline 
Quantum RL \cite{dong2008quantum} & Gate model & - & - & Train, Use Policy \\ \hline 
\end{tabular}
\caption{Selection of QAI algorithms.}
\label{table:algorithm_table}
\vspace{-3em}
\end{table*}

\section{Challenges of Quantum-Assisted Artificial Intelligence}
\label{sec:chall}

Having analyzed the needs of AI and the current state of QAI, we can use this background knowledge and derive major challenges for future developments in QAI. 
Note that unlike other work~\cite{manju2014applications,perdomo2018opportunities} that formulates challenges in quantum artificial intelligence, we focus less on quantum-technical challenges but on the changes to the development methods that need to be achieved.

\begin{chall}[The Feedback Loop]
	Replace the feedback loop around training (consisting of the tasks ``Select Model/Policy'', ``Train'', and ``Assess QoS'') entirely with a quantum algorithm.
\end{chall}

When performing machine learning, a lot of time is usually spent in training, which usually means fine-tuning a set of parameters in small gradual steps over many iterations. These iterations are often necessary as they incorporate slightly different (sets of) data points into the final model. Here, quantum approaches might not treat training iterations as a sequence of steps but maybe perform all training iterations in superposition und thus taking a huge shortcut in training a machine learning model. However, none of the surveyed approaches managed to replace such large parts of the machine learning pipeline by quantum approaches, perhaps because real(istic) quantum machines only provide relatively small coherence times. Quantum RL~\cite{dong2008quantum} probably comes closest by performing both the action execution and the resulting update in single run on the quantum machine, but the algorithm still requires many iterations of training overall. If possible at all, stepping away from iterative training might be the single biggest performance increase quantum computing could offer for AI. Thus, we might refer to The Feedback Loop Challenge as the ``Holy Grail of Quantum AI''.

Nonetheless, other challenges persist and might be detrimental to achieving this highest of goals. Considering the multitude of QAI algorithms focusing on the domain model, we see that quantum-based representations can be used as models for physical domains (where they are a natural fit), complex stochastic domain (where they can approximate complex probability distributions cheaper and more precisely) and small domains in general (where quantum-based or quantum-assisted modeling of the domain might yield some benefits further along the pipeline).

\begin{figure}
	\centering
	\includegraphics[width=.6\columnwidth]{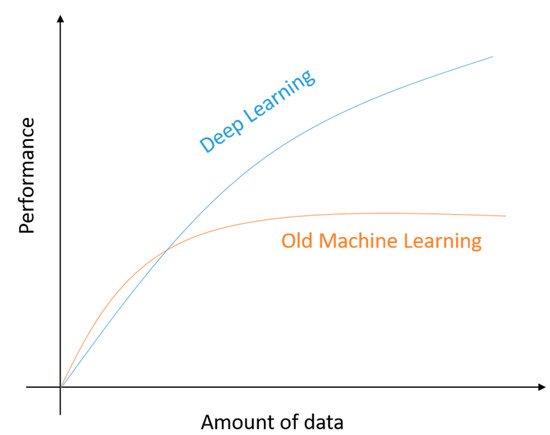}
	\caption{The performance of modern deep learning methods compared to more traditional machine learning depending on the amount of available training data (image taken from~\cite{alom2019state}).}
	\label{fig:data}
\end{figure}

That the extremely limited memory capacities of current quantum computers are one of the main bottlenecks for practical applications is well-known among the quantum computing. However, for QAI algorithms, especially the more modern ones, this problem is aggravated as only through processing very large amounts of data modern AI algorithms really shine~\cite{alom2019state}. Figure~\ref{fig:data} shows a simple sketch of that behavior. Effectively, the need to process relatively large amounts of training data might even, in the long run, prevent us from cutting out the iterative training loop.

\begin{chall}[The Training Data]
	Provide means to process (the essence of) large amounts of data on quantum computers.
\end{chall}

Note that for QAI, we might take a workaround here: Using the right hybrid approaches we might be able to construct classical pre-/postprocessing steps so that we can still process large amounts of data without processing all of them on the quantum machine. Early approaches like Quantum-enhanced RL~\cite{neukart2018quantum} have improved classical training by doing a preselection of training samples (using a quantum algorithm). Similar approaches could work to reduce the necessary training data for quantum training steps as well.

From these considerations we can already see that the combination and hybridization of various algorithms and techniques might be key to further developing QAI. However, combinations always include additional free parameters: What algorithms do we use? How and when do they interact? What domains is a specific combination good for? Furthermore, we do not only need to combine different techniques, but these techniques often stem from different fields of science and engineering. That means that even for a relatively standard QAI algorithm, we might require expert knowledge about quantum computing and the platforms it is run on, about AI and classical optimization, and about the domain at hand in order to make the right calls.

\begin{chall}[The Interfaces]
	Provide standardized interfaces that allow for dynamic combination of QAI components and (by extension) for experts of different fields to collaborate on QAI algorithms. 
\end{chall}

Standardization is a goal that is often called for throughout various disciplines of science and engineering. However, QAI brings together two largely separate field, which in their own right develop rapidly and have produced little standardization. It thus be imperative to organize the interfaces between AI and QC without fixed technological standards but based on the involved experts of different expertise~\cite{leymann2019towards}. An important part of this challenge is to allow standard software engineering to catch up with recent developments: Especially smaller groups will not be able to afford dedicated experts in QC and much less QAI. Instead software developers should be able to use QAI as seamlessly as they are able to use parallel computing in the cloud now, being able to benefit from advantages without the need to dive into the technical specifics.

For QC, this challenge requires a degree of technical maturity that is as of yet not reached by most practical frameworks, even though recent developments definitely aim towards making QC technology more accessible. As a lot of effort is put into QC by vendors wanting to sell their applications, the independent development of open standards is required to prevent vendor lock-in and enable QAI applications that span different QC platforms.


\begin{chall}[The Real Reason]
	Keep track of the source of observed improvements.
\end{chall}

Even classical machine learning models can often be treated as nothing more than a black box; even though they are deterministic and mathematically well understood, they just encode a behavior or connections between input and output that are too complex to trace without extreme computational effort. This is why in recent years, AI researchers showed increased interest in methods of testing and verifying the performance of AI~\cite{calinescu2012self,belzner2016software,amodei2016concrete}.

For QAI, this black box property may be enforced by nature: We physically cannot introspect the probability distribution of states of a quantum machine while it is computing. That is all the more reason why we need quantum-appropriate testing and verification. Under this light, it is rather curious that we found no QAI algorithms that specifically tackle the last few tasks of the machine learning pipeline, especially ``Monitor QoS'', which should be of utmost importance to practical applications.

Challenge~4, however, focuses on the reason why we need especially thorough testing in QAI: We need to constantly justify using a quantum machine. QAI will only have a success if the quantum part of the algorithm is the part that brings about the advantage over comparable methods. However, especially in the field of AI it is easy to construct a superior AI model \emph{by accident}: A few lucky random numbers in the stochastic training process  might result in a better performing AI. Or any part of a QAI algorithm (made up of various classical parts as well) might just match the current (state of the) domain the right way. 

The more complex QAI algorithms become, the harder it might be to find a fair comparison in the purely classical world. Still, we need to provide researchers and developers in the field of QAI with the right tools to easily trace the  significance and the reason of perceived advantages in comparison to other algorithms. If QC is eventually going to benefit AI, we need to be able to know exactly when and for what reason.

\section{Outlook}
\label{sec:outlook}

In this paper we took a long tour from the challenges AI already poses to software engineering to the even more peculiar challenges that QAI poses to software engineering. Still, we argued that QC may greatly help in alleviating the problems the development of increasingly better AI is going to face in the upcoming years. On the flip side, AI methods with inherent robustness to noise might be an ideal testbed for early NISQ applications.

We defined four major challenges that stand without any claim to completeness. On the contrary, we expect every researcher in the field to be able to add quite a few more. However, we feel that the analysis of the projected future developments in AI and the current state of the art in QAI allowed us to deduce some of the most ambitious goals to tackle.

We hope that discussing these highly aimed challenges benefits the development of the young field of QAI and are confident that future research will (purposefully or inadvertently) make progress with respect to these  challenges.

\begin{acks}
This work was supported by the Federal Ministry of Economic Affairs and Energy, Germany, as part of the PlanQK project developing a platform and ecosystem for quantum-assisted artificial intelligence (see \url{planqk.de}).
\end{acks}

\bibliographystyle{ACM-Reference-Format}
\bibliography{references,general,algorithms}


\begin{thebibliography}{56}


\ifx \showCODEN    \undefined \def \showCODEN     #1{\unskip}     \fi
\ifx \showDOI      \undefined \def \showDOI       #1{#1}\fi
\ifx \showISBNx    \undefined \def \showISBNx     #1{\unskip}     \fi
\ifx \showISBNxiii \undefined \def \showISBNxiii  #1{\unskip}     \fi
\ifx \showISSN     \undefined \def \showISSN      #1{\unskip}     \fi
\ifx \showLCCN     \undefined \def \showLCCN      #1{\unskip}     \fi
\ifx \shownote     \undefined \def \shownote      #1{#1}          \fi
\ifx \showarticletitle \undefined \def \showarticletitle #1{#1}   \fi
\ifx \showURL      \undefined \def \showURL       {\relax}        \fi
\providecommand\bibfield[2]{#2}
\providecommand\bibinfo[2]{#2}
\providecommand\natexlab[1]{#1}
\providecommand\showeprint[2][]{arXiv:#2}

\bibitem[\protect\citeauthoryear{??}{hhl}{[n.d.]}]%
        {hhl_qiskit}
 \bibinfo{year}{[n.d.]}\natexlab{}.
\newblock \bibinfo{title}{HHL implementation}.
\newblock
\newblock
\newblock
\shownote{\url{https://github.com/Qiskit/qiskit-iqx-tutorials/blob/master/qiskit/advanced/aqua/linear_systems_of_equations.ipynb}.}


\bibitem[\protect\citeauthoryear{??}{qao}{[n.d.]}]%
        {qaoa_pennylane}
 \bibinfo{year}{[n.d.]}\natexlab{}.
\newblock \bibinfo{title}{QAOA implementation}.
\newblock
\newblock
\newblock
\shownote{\url{https://pennylane.ai/qml/app/tutorial_qaoa_maxcut.html}.}


\bibitem[\protect\citeauthoryear{??}{qga}{[n.d.]a}]%
        {qgan_pennylane}
 \bibinfo{year}{[n.d.]}\natexlab{a}.
\newblock \bibinfo{title}{QGAN Pennylane implementation}.
\newblock
\newblock
\newblock
\shownote{\url{https://pennylane.ai/qml/app/tutorial_QGAN.html}.}


\bibitem[\protect\citeauthoryear{??}{qga}{[n.d.]b}]%
        {qgan_qiskit}
 \bibinfo{year}{[n.d.]}\natexlab{b}.
\newblock \bibinfo{title}{QGAN Qiskit implementation}.
\newblock
\newblock
\newblock
\shownote{\url{https://github.com/Qiskit/qiskit-iqx-tutorials/blob/master/qiskit/advanced/aqua/artificial_intelligence/qgans_for_loading_random_distributions.ipynb}.}


\bibitem[\protect\citeauthoryear{??}{svm}{[n.d.]}]%
        {svm_qiskit}
 \bibinfo{year}{[n.d.]}\natexlab{}.
\newblock \bibinfo{title}{SVM implementation}.
\newblock
\newblock
\newblock
\shownote{\url{https://github.com/Qiskit/qiskit-iqx-tutorials/blob/master/qiskit/advanced/aqua/artificial_intelligence/qsvm_classification.ipynb}.}


\bibitem[\protect\citeauthoryear{A{\"\i}meur, Brassard, and Gambs}{A{\"\i}meur
  et~al\mbox{.}}{2007}]%
        {aimeur2007quantum}
\bibfield{author}{\bibinfo{person}{Esma A{\"\i}meur}, \bibinfo{person}{Gilles
  Brassard}, {and} \bibinfo{person}{S{\'e}bastien Gambs}.}
  \bibinfo{year}{2007}\natexlab{}.
\newblock \showarticletitle{Quantum clustering algorithms}.
\newblock \bibinfo{journal}{\emph{Proceedings of the 24th int'l conference on
  machine learning}} (\bibinfo{year}{2007}).
\newblock


\bibitem[\protect\citeauthoryear{Albers, Sprott, and Dechert}{Albers
  et~al\mbox{.}}{1996}]%
        {albers1996dynamical}
\bibfield{author}{\bibinfo{person}{DJ Albers}, \bibinfo{person}{JC Sprott},
  {and} \bibinfo{person}{WD Dechert}.} \bibinfo{year}{1996}\natexlab{}.
\newblock \showarticletitle{Dynamical behavior of artificial neural networks
  with random weights}.
\newblock \bibinfo{journal}{\emph{Intelligent Engineering Systems Through
  Artificial Neural Networks}}  \bibinfo{volume}{6} (\bibinfo{year}{1996}),
  \bibinfo{pages}{17--22}.
\newblock


\bibitem[\protect\citeauthoryear{Alom, Taha, Yakopcic, Westberg, Sidike,
  Nasrin, Hasan, Van~Essen, Awwal, and Asari}{Alom et~al\mbox{.}}{2019}]%
        {alom2019state}
\bibfield{author}{\bibinfo{person}{Md~Zahangir Alom}, \bibinfo{person}{Tarek~M
  Taha}, \bibinfo{person}{Chris Yakopcic}, \bibinfo{person}{Stefan Westberg},
  \bibinfo{person}{Paheding Sidike}, \bibinfo{person}{Mst~Shamima Nasrin},
  \bibinfo{person}{Mahmudul Hasan}, \bibinfo{person}{Brian~C Van~Essen},
  \bibinfo{person}{Abdul~AS Awwal}, {and} \bibinfo{person}{Vijayan~K Asari}.}
  \bibinfo{year}{2019}\natexlab{}.
\newblock \showarticletitle{A state-of-the-art survey on deep learning theory
  and architectures}.
\newblock \bibinfo{journal}{\emph{Electronics}} \bibinfo{volume}{8},
  \bibinfo{number}{3} (\bibinfo{year}{2019}), \bibinfo{pages}{292}.
\newblock


\bibitem[\protect\citeauthoryear{Amin, Andriyash, Rolfe, Kulchytskyy, and
  Melko}{Amin et~al\mbox{.}}{2018}]%
        {amin2018quantum}
\bibfield{author}{\bibinfo{person}{Mohammad~H Amin}, \bibinfo{person}{Evgeny
  Andriyash}, \bibinfo{person}{Jason Rolfe}, \bibinfo{person}{Bohdan
  Kulchytskyy}, {and} \bibinfo{person}{Roger Melko}.}
  \bibinfo{year}{2018}\natexlab{}.
\newblock \showarticletitle{Quantum boltzmann machine}.
\newblock \bibinfo{journal}{\emph{arXiv:1601.02036}} (\bibinfo{year}{2018}).
\newblock


\bibitem[\protect\citeauthoryear{Amodei and Hernandez}{Amodei and
  Hernandez}{2018}]%
        {openai2018compute}
\bibfield{author}{\bibinfo{person}{Dario Amodei} {and} \bibinfo{person}{Danny
  Hernandez}.} \bibinfo{year}{2018}\natexlab{}.
\newblock \bibinfo{title}{{AI} and Compute}.
\newblock
  \bibinfo{howpublished}{\url{https://openai.com/blog/ai-and-compute/}}.
\newblock


\bibitem[\protect\citeauthoryear{Amodei, Olah, Steinhardt, Christiano,
  Schulman, and Man{\'e}}{Amodei et~al\mbox{.}}{2016}]%
        {amodei2016concrete}
\bibfield{author}{\bibinfo{person}{Dario Amodei}, \bibinfo{person}{Chris Olah},
  \bibinfo{person}{Jacob Steinhardt}, \bibinfo{person}{Paul Christiano},
  \bibinfo{person}{John Schulman}, {and} \bibinfo{person}{Dan Man{\'e}}.}
  \bibinfo{year}{2016}\natexlab{}.
\newblock \showarticletitle{Concrete problems in {AI} safety}.
\newblock \bibinfo{journal}{\emph{arXiv:1606.06565}} (\bibinfo{year}{2016}).
\newblock


\bibitem[\protect\citeauthoryear{Arute, Arya, Babbush, Bacon, Bardin, Barends,
  Biswas, Boixo, Brandao, Buell, et~al\mbox{.}}{Arute et~al\mbox{.}}{2019}]%
        {arute2019quantum}
\bibfield{author}{\bibinfo{person}{Frank Arute}, \bibinfo{person}{Kunal Arya},
  \bibinfo{person}{Ryan Babbush}, \bibinfo{person}{Dave Bacon},
  \bibinfo{person}{Joseph~C Bardin}, \bibinfo{person}{Rami Barends},
  \bibinfo{person}{Rupak Biswas}, \bibinfo{person}{Sergio Boixo},
  \bibinfo{person}{Fernando~GSL Brandao}, \bibinfo{person}{David~A Buell},
  {et~al\mbox{.}}} \bibinfo{year}{2019}\natexlab{}.
\newblock \showarticletitle{Quantum supremacy using a programmable
  superconducting processor}.
\newblock \bibinfo{journal}{\emph{Nature}} \bibinfo{volume}{574},
  \bibinfo{number}{7779} (\bibinfo{year}{2019}), \bibinfo{pages}{505--510}.
\newblock


\bibitem[\protect\citeauthoryear{Belzner, Beck, Gabor, Roelle, and
  Sauer}{Belzner et~al\mbox{.}}{2016}]%
        {belzner2016software}
\bibfield{author}{\bibinfo{person}{Lenz Belzner}, \bibinfo{person}{Michael~Till
  Beck}, \bibinfo{person}{Thomas Gabor}, \bibinfo{person}{Harald Roelle}, {and}
  \bibinfo{person}{Horst Sauer}.} \bibinfo{year}{2016}\natexlab{}.
\newblock \showarticletitle{Software engineering for distributed autonomous
  real-time systems}. In \bibinfo{booktitle}{\emph{Proceedings of the 2nd
  International Workshop on Software Engineering for Smart Cyber-Physical
  Systems}}. ACM, \bibinfo{pages}{54--57}.
\newblock


\bibitem[\protect\citeauthoryear{Bures, Weyns, Schmer, Tovar, Boden, Gabor,
  Gerostathopoulos, Gupta, Kang, Knauss, et~al\mbox{.}}{Bures
  et~al\mbox{.}}{2017}]%
        {bures2017software}
\bibfield{author}{\bibinfo{person}{Tomas Bures}, \bibinfo{person}{Danny Weyns},
  \bibinfo{person}{Bradley Schmer}, \bibinfo{person}{Eduardo Tovar},
  \bibinfo{person}{Eric Boden}, \bibinfo{person}{Thomas Gabor},
  \bibinfo{person}{Ilias Gerostathopoulos}, \bibinfo{person}{Pragya Gupta},
  \bibinfo{person}{Eunsuk Kang}, \bibinfo{person}{Alessia Knauss},
  {et~al\mbox{.}}} \bibinfo{year}{2017}\natexlab{}.
\newblock \showarticletitle{Software Engineering for Smart Cyber-Physical
  Systems: Challenges and Promising Solutions}.
\newblock \bibinfo{journal}{\emph{ACM SIGSOFT Software Eng. Notes}}
  \bibinfo{volume}{42}, \bibinfo{number}{2} (\bibinfo{year}{2017}),
  \bibinfo{pages}{19--24}.
\newblock


\bibitem[\protect\citeauthoryear{Calinescu, Ghezzi, Kwiatkowska, and
  Mirandola}{Calinescu et~al\mbox{.}}{2012}]%
        {calinescu2012self}
\bibfield{author}{\bibinfo{person}{Radu Calinescu}, \bibinfo{person}{Carlo
  Ghezzi}, \bibinfo{person}{Marta Kwiatkowska}, {and} \bibinfo{person}{Raffaela
  Mirandola}.} \bibinfo{year}{2012}\natexlab{}.
\newblock \showarticletitle{Self-adaptive software needs quantitative
  verification at runtime}.
\newblock \bibinfo{journal}{\emph{Commun. ACM}} \bibinfo{volume}{55},
  \bibinfo{number}{9} (\bibinfo{year}{2012}), \bibinfo{pages}{69--77}.
\newblock


\bibitem[\protect\citeauthoryear{Cuccu, Togelius, and Cudr{\'e}-Mauroux}{Cuccu
  et~al\mbox{.}}{2019}]%
        {cuccu2019playing}
\bibfield{author}{\bibinfo{person}{Giuseppe Cuccu}, \bibinfo{person}{Julian
  Togelius}, {and} \bibinfo{person}{Philippe Cudr{\'e}-Mauroux}.}
  \bibinfo{year}{2019}\natexlab{}.
\newblock \showarticletitle{Playing atari with six neurons}. In
  \bibinfo{booktitle}{\emph{Proceedings of the 18th international conference on
  autonomous agents and multiagent systems}}. International Foundation for
  Autonomous Agents and Multiagent Systems, \bibinfo{pages}{998--1006}.
\newblock


\bibitem[\protect\citeauthoryear{Dallaire-Demers and Killoran}{Dallaire-Demers
  and Killoran}{2018}]%
        {dallaire2018quantum}
\bibfield{author}{\bibinfo{person}{Pierre-Luc Dallaire-Demers} {and}
  \bibinfo{person}{Nathan Killoran}.} \bibinfo{year}{2018}\natexlab{}.
\newblock \showarticletitle{Quantum generative adversarial networks}.
\newblock \bibinfo{journal}{\emph{arXiv:1804.08641v2}} (\bibinfo{year}{2018}).
\newblock


\bibitem[\protect\citeauthoryear{Dong, Chen, Li, and Tarn}{Dong
  et~al\mbox{.}}{2008}]%
        {dong2008quantum}
\bibfield{author}{\bibinfo{person}{Daoyi Dong}, \bibinfo{person}{Chunlin Chen},
  \bibinfo{person}{Hanxiong Li}, {and} \bibinfo{person}{Tzyh-Jong Tarn}.}
  \bibinfo{year}{2008}\natexlab{}.
\newblock \showarticletitle{Quantum reinforcement learning}.
\newblock \bibinfo{journal}{\emph{arXiv:0810.3828}} (\bibinfo{year}{2008}).
\newblock


\bibitem[\protect\citeauthoryear{Farhi, Goldstone, and Gutmann}{Farhi
  et~al\mbox{.}}{2014}]%
        {farhi2014quantum}
\bibfield{author}{\bibinfo{person}{Edward Farhi}, \bibinfo{person}{Jeffrey
  Goldstone}, {and} \bibinfo{person}{Sam Gutmann}.}
  \bibinfo{year}{2014}\natexlab{}.
\newblock \showarticletitle{A quantum approximate optimization algorithm}.
\newblock \bibinfo{journal}{\emph{arXiv:1411.4028}} (\bibinfo{year}{2014}).
\newblock


\bibitem[\protect\citeauthoryear{F{\"o}sel, Tighineanu, Weiss, and
  Marquardt}{F{\"o}sel et~al\mbox{.}}{2018}]%
        {fosel2018reinforcement}
\bibfield{author}{\bibinfo{person}{Thomas F{\"o}sel}, \bibinfo{person}{Petru
  Tighineanu}, \bibinfo{person}{Talitha Weiss}, {and} \bibinfo{person}{Florian
  Marquardt}.} \bibinfo{year}{2018}\natexlab{}.
\newblock \showarticletitle{Reinforcement learning with neural networks for
  quantum feedback}.
\newblock \bibinfo{journal}{\emph{Physical Review X}} \bibinfo{volume}{8},
  \bibinfo{number}{3} (\bibinfo{year}{2018}), \bibinfo{pages}{031084}.
\newblock


\bibitem[\protect\citeauthoryear{Gabor, Kiermeier, Sedlmeier, Kempter, Klein,
  Sauer, Schmid, and Wieghardt}{Gabor et~al\mbox{.}}{2018}]%
        {gabor2018adapting}
\bibfield{author}{\bibinfo{person}{Thomas Gabor}, \bibinfo{person}{Marie
  Kiermeier}, \bibinfo{person}{Andreas Sedlmeier}, \bibinfo{person}{Bernhard
  Kempter}, \bibinfo{person}{Cornel Klein}, \bibinfo{person}{Horst Sauer},
  \bibinfo{person}{Reiner Schmid}, {and} \bibinfo{person}{Jan Wieghardt}.}
  \bibinfo{year}{2018}\natexlab{}.
\newblock \showarticletitle{Adapting quality assurance to adaptive systems: the
  scenario coevolution paradigm}. In \bibinfo{booktitle}{\emph{International
  Symposium on Leveraging Applications of Formal Methods}}. Springer,
  \bibinfo{pages}{137--154}.
\newblock


\bibitem[\protect\citeauthoryear{Gabor, Sedlmeier, Phan, Ritz, Kiermeier,
  Belzner, Kempter, Klein, Sauer, Schmid, Wieghardt, Zeller, and
  Linnhoff-Popien}{Gabor et~al\mbox{.}}{2020}]%
        {reocas2020}
\bibfield{author}{\bibinfo{person}{Thomas Gabor}, \bibinfo{person}{Andreas
  Sedlmeier}, \bibinfo{person}{Thomy Phan}, \bibinfo{person}{Fabian Ritz},
  \bibinfo{person}{Marie Kiermeier}, \bibinfo{person}{Lenz Belzner},
  \bibinfo{person}{Bernhard Kempter}, \bibinfo{person}{Cornel Klein},
  \bibinfo{person}{Horst Sauer}, \bibinfo{person}{Reiner Schmid},
  \bibinfo{person}{Jan Wieghardt}, \bibinfo{person}{Marc Zeller}, {and}
  \bibinfo{person}{Claudia Linnhoff-Popien}.} \bibinfo{year}{2020}\natexlab{}.
\newblock \showarticletitle{The Scenario Co-Evolution Paradigm: Adaptive
  Quality Assurance for Adaptive Systems}.
\newblock \bibinfo{journal}{\emph{International Journal on Software Tools and
  Technology Transfer}} (\bibinfo{year}{2020}).
\newblock


\bibitem[\protect\citeauthoryear{Glover, Kochenberger, and Du}{Glover
  et~al\mbox{.}}{2018}]%
        {glover2018tutorial}
\bibfield{author}{\bibinfo{person}{Fred Glover}, \bibinfo{person}{Gary
  Kochenberger}, {and} \bibinfo{person}{Yu Du}.}
  \bibinfo{year}{2018}\natexlab{}.
\newblock \showarticletitle{A tutorial on formulating and using qubo models}.
\newblock \bibinfo{journal}{\emph{arXiv preprint arXiv:1811.11538}}
  (\bibinfo{year}{2018}).
\newblock


\bibitem[\protect\citeauthoryear{Harrow, Hassidim, and Lloyd}{Harrow
  et~al\mbox{.}}{2009}]%
        {harrow2009quantum}
\bibfield{author}{\bibinfo{person}{Aram~W Harrow}, \bibinfo{person}{Avinatan
  Hassidim}, {and} \bibinfo{person}{Seth Lloyd}.}
  \bibinfo{year}{2009}\natexlab{}.
\newblock \showarticletitle{Quantum algorithm for linear systems of equations}.
\newblock \bibinfo{journal}{\emph{arXiv:0811.317v3}} (\bibinfo{year}{2009}).
\newblock


\bibitem[\protect\citeauthoryear{Havlicek, C{\'o}rcoles, Temme, Harrow,
  Kandala, Chow, and Gambetta}{Havlicek et~al\mbox{.}}{2018}]%
        {havlivcek2018supervised}
\bibfield{author}{\bibinfo{person}{Vojt{\v{e}}ch Havlicek},
  \bibinfo{person}{Antonio~D C{\'o}rcoles}, \bibinfo{person}{Kristan Temme},
  \bibinfo{person}{Aram~W Harrow}, \bibinfo{person}{Abhinav Kandala},
  \bibinfo{person}{Jerry~M Chow}, {and} \bibinfo{person}{Jay~M Gambetta}.}
  \bibinfo{year}{2018}\natexlab{}.
\newblock \showarticletitle{Supervised learning with quantum-enhanced feature
  spaces}.
\newblock \bibinfo{journal}{\emph{arXiv:1804.11326v2}} (\bibinfo{year}{2018}).
\newblock


\bibitem[\protect\citeauthoryear{H{\"o}lzl, Koch, Puviani, Wirsing, and
  Zambonelli}{H{\"o}lzl et~al\mbox{.}}{2015}]%
        {holzl2015ensemble}
\bibfield{author}{\bibinfo{person}{Matthias H{\"o}lzl}, \bibinfo{person}{Nora
  Koch}, \bibinfo{person}{Mariachiara Puviani}, \bibinfo{person}{Martin
  Wirsing}, {and} \bibinfo{person}{Franco Zambonelli}.}
  \bibinfo{year}{2015}\natexlab{}.
\newblock \showarticletitle{The ensemble development life cycle and best
  practices for collective autonomic systems}.
\newblock In \bibinfo{booktitle}{\emph{Software Engineering for Collective
  Autonomic Systems}}. \bibinfo{publisher}{Springer},
  \bibinfo{pages}{325--354}.
\newblock


\bibitem[\protect\citeauthoryear{Kadowaki and Nishimori}{Kadowaki and
  Nishimori}{1998}]%
        {kadowaki1998quantum}
\bibfield{author}{\bibinfo{person}{Tadashi Kadowaki} {and}
  \bibinfo{person}{Hidetoshi Nishimori}.} \bibinfo{year}{1998}\natexlab{}.
\newblock \showarticletitle{Quantum annealing in the transverse Ising model}.
\newblock \bibinfo{journal}{\emph{Physical Review E}} \bibinfo{volume}{58},
  \bibinfo{number}{5} (\bibinfo{year}{1998}), \bibinfo{pages}{5355}.
\newblock


\bibitem[\protect\citeauthoryear{Kerenidis and Prakash}{Kerenidis and
  Prakash}{2016}]%
        {kerenidis2016quantum}
\bibfield{author}{\bibinfo{person}{Iordanis Kerenidis} {and}
  \bibinfo{person}{Anupam Prakash}.} \bibinfo{year}{2016}\natexlab{}.
\newblock \showarticletitle{Quantum recommendation systems}.
\newblock \bibinfo{journal}{\emph{arXiv:1603.08675v3}} (\bibinfo{year}{2016}).
\newblock


\bibitem[\protect\citeauthoryear{Khoshaman, Vinci, Denis, Andriyash, Sadeghi,
  and Amin}{Khoshaman et~al\mbox{.}}{2018}]%
        {khoshaman2018quantum}
\bibfield{author}{\bibinfo{person}{Amir Khoshaman}, \bibinfo{person}{Walter
  Vinci}, \bibinfo{person}{Brandon Denis}, \bibinfo{person}{Evgeny Andriyash},
  \bibinfo{person}{Hossein Sadeghi}, {and} \bibinfo{person}{Mohammad~H Amin}.}
  \bibinfo{year}{2018}\natexlab{}.
\newblock \showarticletitle{Quantum variational autoencoder}.
\newblock \bibinfo{journal}{\emph{arXiv:1802.05779v2}} (\bibinfo{year}{2018}).
\newblock


\bibitem[\protect\citeauthoryear{King, Mohseni, Bernoudy, Fr{\'e}chette,
  Sadeghi, Isakov, Neven, and Amin}{King et~al\mbox{.}}{2019}]%
        {king2019quantum}
\bibfield{author}{\bibinfo{person}{James King}, \bibinfo{person}{Masoud
  Mohseni}, \bibinfo{person}{William Bernoudy}, \bibinfo{person}{Alexandre
  Fr{\'e}chette}, \bibinfo{person}{Hossein Sadeghi}, \bibinfo{person}{Sergei~V
  Isakov}, \bibinfo{person}{Hartmut Neven}, {and} \bibinfo{person}{Mohammad~H
  Amin}.} \bibinfo{year}{2019}\natexlab{}.
\newblock \showarticletitle{Quantum-Assisted Genetic Algorithm}.
\newblock \bibinfo{journal}{\emph{arXiv:1907.00707}} (\bibinfo{year}{2019}).
\newblock


\bibitem[\protect\citeauthoryear{Leymann, Barzen, and Falkenthal}{Leymann
  et~al\mbox{.}}{2019}]%
        {leymann2019towards}
\bibfield{author}{\bibinfo{person}{Frank Leymann}, \bibinfo{person}{Johanna
  Barzen}, {and} \bibinfo{person}{Michael Falkenthal}.}
  \bibinfo{year}{2019}\natexlab{}.
\newblock \showarticletitle{Towards a Platform for Sharing Quantum Software}.
\newblock \bibinfo{journal}{\emph{Proceedings of the 13th Advanced Summer
  School on Service}} (\bibinfo{year}{2019}), \bibinfo{pages}{70--74}.
\newblock


\bibitem[\protect\citeauthoryear{Lloyd, Mohseni, and Rebentrost}{Lloyd
  et~al\mbox{.}}{2013}]%
        {lloyd2013quantum}
\bibfield{author}{\bibinfo{person}{Seth Lloyd}, \bibinfo{person}{Masoud
  Mohseni}, {and} \bibinfo{person}{Patrick Rebentrost}.}
  \bibinfo{year}{2013}\natexlab{}.
\newblock \showarticletitle{Quantum algorithms for supervised and unsupervised
  machine learning}.
\newblock \bibinfo{journal}{\emph{arXiv:1307.0411}} (\bibinfo{year}{2013}).
\newblock


\bibitem[\protect\citeauthoryear{Lloyd and Weedbrook}{Lloyd and
  Weedbrook}{2018}]%
        {lloyd2018quantum}
\bibfield{author}{\bibinfo{person}{Seth Lloyd} {and} \bibinfo{person}{Christian
  Weedbrook}.} \bibinfo{year}{2018}\natexlab{}.
\newblock \showarticletitle{Quantum generative adversarial learning}.
\newblock \bibinfo{journal}{\emph{arXiv:1804.09139}} (\bibinfo{year}{2018}).
\newblock


\bibitem[\protect\citeauthoryear{Lucas}{Lucas}{2014}]%
        {lucas2014ising}
\bibfield{author}{\bibinfo{person}{Andrew Lucas}.}
  \bibinfo{year}{2014}\natexlab{}.
\newblock \showarticletitle{Ising formulations of many NP problems}.
\newblock \bibinfo{journal}{\emph{Frontiers in Physics}}  \bibinfo{volume}{2}
  (\bibinfo{year}{2014}), \bibinfo{pages}{5}.
\newblock


\bibitem[\protect\citeauthoryear{Manju and Nigam}{Manju and Nigam}{2014}]%
        {manju2014applications}
\bibfield{author}{\bibinfo{person}{A Manju} {and} \bibinfo{person}{Madhav~J
  Nigam}.} \bibinfo{year}{2014}\natexlab{}.
\newblock \showarticletitle{Applications of quantum inspired computational
  intelligence: a survey}.
\newblock \bibinfo{journal}{\emph{Artificial Intelligence Review}}
  \bibinfo{volume}{42}, \bibinfo{number}{1} (\bibinfo{year}{2014}),
  \bibinfo{pages}{79--156}.
\newblock


\bibitem[\protect\citeauthoryear{McGeoch}{McGeoch}{2014}]%
        {mcgeoch2014adiabatic}
\bibfield{author}{\bibinfo{person}{Catherine~C McGeoch}.}
  \bibinfo{year}{2014}\natexlab{}.
\newblock \showarticletitle{Adiabatic quantum computation and quantum
  annealing: Theory and practice}.
\newblock \bibinfo{journal}{\emph{Synthesis Lectures on Quantum Computing}}
  \bibinfo{volume}{5}, \bibinfo{number}{2} (\bibinfo{year}{2014}),
  \bibinfo{pages}{1--93}.
\newblock


\bibitem[\protect\citeauthoryear{McGeoch and Wang}{McGeoch and Wang}{2013}]%
        {mcgeoch2013experimental}
\bibfield{author}{\bibinfo{person}{Catherine~C McGeoch} {and}
  \bibinfo{person}{Cong Wang}.} \bibinfo{year}{2013}\natexlab{}.
\newblock \showarticletitle{Experimental evaluation of an adiabiatic quantum
  system for combinatorial optimization}. In
  \bibinfo{booktitle}{\emph{Proceedings of the ACM International Conference on
  Computing Frontiers}}. \bibinfo{pages}{1--11}.
\newblock


\bibitem[\protect\citeauthoryear{Nachum, Gu, Lee, and Levine}{Nachum
  et~al\mbox{.}}{2018}]%
        {nachum2018data}
\bibfield{author}{\bibinfo{person}{Ofir Nachum},
  \bibinfo{person}{Shixiang~Shane Gu}, \bibinfo{person}{Honglak Lee}, {and}
  \bibinfo{person}{Sergey Levine}.} \bibinfo{year}{2018}\natexlab{}.
\newblock \showarticletitle{Data-efficient hierarchical reinforcement
  learning}. In \bibinfo{booktitle}{\emph{Advances in Neural Information
  Processing Systems}}. \bibinfo{pages}{3303--3313}.
\newblock


\bibitem[\protect\citeauthoryear{Neukart, Von~Dollen, Seidel, and
  Compostella}{Neukart et~al\mbox{.}}{2018}]%
        {neukart2018quantum}
\bibfield{author}{\bibinfo{person}{Florian Neukart}, \bibinfo{person}{David
  Von~Dollen}, \bibinfo{person}{Christian Seidel}, {and}
  \bibinfo{person}{Gabriele Compostella}.} \bibinfo{year}{2018}\natexlab{}.
\newblock \showarticletitle{Quantum-enhanced reinforcement learning for
  finite-episode games with discrete state spaces}.
\newblock \bibinfo{journal}{\emph{arXiv:1708.09354v3}} (\bibinfo{year}{2018}).
\newblock


\bibitem[\protect\citeauthoryear{Nierstrasz, Denker, G{\^\i}rba, Lienhard, and
  R{\"o}thlisberger}{Nierstrasz et~al\mbox{.}}{2008}]%
        {nierstrasz2008change}
\bibfield{author}{\bibinfo{person}{Oscar Nierstrasz}, \bibinfo{person}{Marcus
  Denker}, \bibinfo{person}{Tudor G{\^\i}rba}, \bibinfo{person}{Adrian
  Lienhard}, {and} \bibinfo{person}{David R{\"o}thlisberger}.}
  \bibinfo{year}{2008}\natexlab{}.
\newblock \showarticletitle{Change-enabled software systems}.
\newblock In \bibinfo{booktitle}{\emph{Software-Intensive Systems and New
  Computing Paradigms}}. \bibinfo{publisher}{Springer},
  \bibinfo{pages}{64--79}.
\newblock


\bibitem[\protect\citeauthoryear{PennyLane}{PennyLane}{[n.d.]}]%
        {vqa_pennylane}
\bibfield{author}{\bibinfo{person}{PennyLane}.}
  \bibinfo{year}{[n.d.]}\natexlab{}.
\newblock \bibinfo{title}{VQE implementation}.
\newblock
\newblock
\newblock
\shownote{\url{https://pennylane.ai/qml/demos/tutorial_vqe.html}.}


\bibitem[\protect\citeauthoryear{Perdomo-Ortiz, Benedetti, Realpe-G{\'o}mez,
  and Biswas}{Perdomo-Ortiz et~al\mbox{.}}{2018}]%
        {perdomo2018opportunities}
\bibfield{author}{\bibinfo{person}{Alejandro Perdomo-Ortiz},
  \bibinfo{person}{Marcello Benedetti}, \bibinfo{person}{John
  Realpe-G{\'o}mez}, {and} \bibinfo{person}{Rupak Biswas}.}
  \bibinfo{year}{2018}\natexlab{}.
\newblock \showarticletitle{Opportunities and challenges for quantum-assisted
  machine learning in near-term quantum computers}.
\newblock \bibinfo{journal}{\emph{Quantum Science and Technology}}
  \bibinfo{volume}{3}, \bibinfo{number}{3} (\bibinfo{year}{2018}),
  \bibinfo{pages}{030502}.
\newblock


\bibitem[\protect\citeauthoryear{Peruzzo, McClean, Shadbolt, Yung, Zhou, Love,
  Aspuru-Guzik, and O’brien}{Peruzzo et~al\mbox{.}}{2014}]%
        {peruzzo2014variational}
\bibfield{author}{\bibinfo{person}{Alberto Peruzzo}, \bibinfo{person}{Jarrod
  McClean}, \bibinfo{person}{Peter Shadbolt}, \bibinfo{person}{Man-Hong Yung},
  \bibinfo{person}{Xiao-Qi Zhou}, \bibinfo{person}{Peter~J Love},
  \bibinfo{person}{Al{\'a}n Aspuru-Guzik}, {and} \bibinfo{person}{Jeremy~L
  O’brien}.} \bibinfo{year}{2014}\natexlab{}.
\newblock \showarticletitle{A variational eigenvalue solver on a photonic
  quantum processor}.
\newblock \bibinfo{journal}{\emph{Nature commun.}}  \bibinfo{volume}{5}
  (\bibinfo{year}{2014}).
\newblock


\bibitem[\protect\citeauthoryear{Porotti, Tamascelli, Restelli, and
  Prati}{Porotti et~al\mbox{.}}{2019}]%
        {porotti2019coherent}
\bibfield{author}{\bibinfo{person}{Riccardo Porotti}, \bibinfo{person}{Dario
  Tamascelli}, \bibinfo{person}{Marcello Restelli}, {and}
  \bibinfo{person}{Enrico Prati}.} \bibinfo{year}{2019}\natexlab{}.
\newblock \showarticletitle{Coherent transport of quantum states by deep
  reinforcement learning}.
\newblock \bibinfo{journal}{\emph{Communications Physics}} \bibinfo{volume}{2},
  \bibinfo{number}{1} (\bibinfo{year}{2019}), \bibinfo{pages}{1--9}.
\newblock


\bibitem[\protect\citeauthoryear{Preskill}{Preskill}{2018}]%
        {preskill2018quantum}
\bibfield{author}{\bibinfo{person}{John Preskill}.}
  \bibinfo{year}{2018}\natexlab{}.
\newblock \showarticletitle{Quantum Computing in the NISQ era and beyond}.
\newblock \bibinfo{journal}{\emph{arXiv:1801.00862v3}} (\bibinfo{year}{2018}).
\newblock


\bibitem[\protect\citeauthoryear{Romero and Aspuru-Guzik}{Romero and
  Aspuru-Guzik}{2019}]%
        {romero2019variational}
\bibfield{author}{\bibinfo{person}{Jonathan Romero} {and} \bibinfo{person}{Alan
  Aspuru-Guzik}.} \bibinfo{year}{2019}\natexlab{}.
\newblock \showarticletitle{Variational quantum generators: Generative
  adversarial quantum machine learning for continuous distributions}.
\newblock \bibinfo{journal}{\emph{arXiv:1901.00848}} (\bibinfo{year}{2019}).
\newblock


\bibitem[\protect\citeauthoryear{Romero, Olson, and Aspuru-Guzik}{Romero
  et~al\mbox{.}}{2017}]%
        {romero2017quantum}
\bibfield{author}{\bibinfo{person}{Jonathan Romero},
  \bibinfo{person}{Jonathan~P Olson}, {and} \bibinfo{person}{Alan
  Aspuru-Guzik}.} \bibinfo{year}{2017}\natexlab{}.
\newblock \showarticletitle{Quantum autoencoders for efficient compression of
  quantum data}.
\newblock \bibinfo{journal}{\emph{arXiv:1612.02806v2}} (\bibinfo{year}{2017}).
\newblock


\bibitem[\protect\citeauthoryear{Sim, Anderson, Brown, and Romero}{Sim
  et~al\mbox{.}}{[n.d.]}]%
        {autoencoder_implementation}
\bibfield{author}{\bibinfo{person}{Sukin Sim}, \bibinfo{person}{Evan Anderson},
  \bibinfo{person}{Eric Brown}, {and} \bibinfo{person}{Jonathan Romero}.}
  \bibinfo{year}{[n.d.]}\natexlab{}.
\newblock \bibinfo{title}{Autoencoder implementation}.
\newblock
\newblock
\newblock
\shownote{\url{https://github.com/hsim13372/QCompress-1}.}


\bibitem[\protect\citeauthoryear{Sutton}{Sutton}{2019}]%
        {sutton2019bitter}
\bibfield{author}{\bibinfo{person}{Richard Sutton}.}
  \bibinfo{year}{2019}\natexlab{}.
\newblock \bibinfo{title}{The Bitter Lesson}.
\newblock
  \bibinfo{howpublished}{\url{http://www.incompleteideas.net/IncIdeas/BitterLesson.html}}.
\newblock


\bibitem[\protect\citeauthoryear{team}{team}{2019}]%
        {alphastar2019}
\bibfield{author}{\bibinfo{person}{The~AlphaStar team}.}
  \bibinfo{year}{2019}\natexlab{}.
\newblock \bibinfo{title}{{AlphaStar}: Mastering the Real-Time Strategy Game
  {StarCraft II}}.
\newblock
  \bibinfo{howpublished}{\url{https://deepmind.com/blog/article/alphastar-mastering-real-time-strategy-game-starcraft-ii}}.
\newblock


\bibitem[\protect\citeauthoryear{Weyns}{Weyns}{2017}]%
        {weyns2017software}
\bibfield{author}{\bibinfo{person}{Danny Weyns}.}
  \bibinfo{year}{2017}\natexlab{}.
\newblock \showarticletitle{Software engineering of self-adaptive systems: an
  organised tour and future challenges}.
\newblock  (\bibinfo{year}{2017}).
\newblock


\bibitem[\protect\citeauthoryear{Wiebe, Kapoor, and Svore}{Wiebe
  et~al\mbox{.}}{2014}]%
        {wiebe2014quantum_algorithms}
\bibfield{author}{\bibinfo{person}{Nathan Wiebe}, \bibinfo{person}{Ashish
  Kapoor}, {and} \bibinfo{person}{Krysta Svore}.}
  \bibinfo{year}{2014}\natexlab{}.
\newblock \showarticletitle{Quantum algorithms for nearest-neighbor methods for
  supervised and unsupervised learning}.
\newblock \bibinfo{journal}{\emph{arXiv:1401.2142v2}} (\bibinfo{year}{2014}).
\newblock


\bibitem[\protect\citeauthoryear{Wiebe and Wossnig}{Wiebe and Wossnig}{2019}]%
        {wiebe2019generative}
\bibfield{author}{\bibinfo{person}{Nathan Wiebe} {and} \bibinfo{person}{Leonard
  Wossnig}.} \bibinfo{year}{2019}\natexlab{}.
\newblock \showarticletitle{Generative training of quantum Boltzmann machines
  with hidden units}.
\newblock \bibinfo{journal}{\emph{arXiv:1905.09902}} (\bibinfo{year}{2019}).
\newblock


\bibitem[\protect\citeauthoryear{Willsch, Willsch, De~Raedt, and
  Michielsen}{Willsch et~al\mbox{.}}{2019}]%
        {willsch2019support}
\bibfield{author}{\bibinfo{person}{D. Willsch}, \bibinfo{person}{M. Willsch},
  \bibinfo{person}{H. De~Raedt}, {and} \bibinfo{person}{K. Michielsen}.}
  \bibinfo{year}{2019}\natexlab{}.
\newblock \showarticletitle{Support vector machines on the D-Wave quantum
  annealer}.
\newblock \bibinfo{journal}{\emph{arXiv:1906.06283}} (\bibinfo{year}{2019}).
\newblock


\bibitem[\protect\citeauthoryear{Wolfram}{Wolfram}{2018}]%
        {wolfram2018making}
\bibfield{author}{\bibinfo{person}{Stephen Wolfram}.}
  \bibinfo{year}{2018}\natexlab{}.
\newblock \bibinfo{title}{Buzzword Convergence: Making Sense of Quantum Neural
  Blockchain {AI}}.
\newblock
  \bibinfo{howpublished}{\url{https://writings.stephenwolfram.com/2018/04/buzzword-convergence-making-sense-of-quantum-neural-blockchain-ai/}}.
\newblock


\bibitem[\protect\citeauthoryear{Zoufal, Lucchi, and Woerner}{Zoufal
  et~al\mbox{.}}{2019}]%
        {zoufal2019quantum}
\bibfield{author}{\bibinfo{person}{Christa Zoufal},
  \bibinfo{person}{Aur{\'e}lien Lucchi}, {and} \bibinfo{person}{Stefan
  Woerner}.} \bibinfo{year}{2019}\natexlab{}.
\newblock \showarticletitle{Quantum generative adversarial networks for
  learning and loading random distributions}.
\newblock \bibinfo{journal}{\emph{arXiv:1904.00043}} (\bibinfo{year}{2019}).
\newblock


\end{thebibliography}
{\footnotesize All URLs have been accessed on March 1, 2020.}

\end{document}